 \documentclass[twocolumn]{aastex6}
\shorttitle{Convectively driven sinks and magnetic fields}
\shortauthors{Requerey et al.}
\begin{document}


\title{Convectively driven sinks and magnetic fields in the quiet Sun}

%% Use \author, \affil, plus the \and command to format author and affiliation 
%% information.  If done correctly the peer review system will be able to
%% automatically put the author and affiliation information from the manuscript
%% and save the corresponding author the trouble of entering it by hand.
%%
%% The \affil should be used to document primary affiliations and the
%% \altaffil should be used for secondary affiliations, titles, or email.

%% Authors with the same affiliation can be grouped in a single
\author{\textsc{
Iker~S.~Requerey$^{1,2,3}$,
Jose~Carlos~Del~Toro~Iniesta$^{1}$, 
Luis~R.~Bellot~Rubio$^{1}$, 
Valent\'{i}n~Mart\'{i}nez~Pillet$^{4}$, 
Sami~K.~Solanki$^{5,6}$,
and Wolfgang~Schmidt$^{7}$
 }}
\affil{
$^{1}$Instituto de Astrof\'{i}sica de Andaluc\'{i}a (CSIC), Apdo. de Correos 3004, E-18080 Granada, Spain\\
$^{2}$Instituto de Astrof\'{i}sica de Canarias, V\'{i}a L\'{a}ctea s/n, E-38205 La Laguna, Tenerife, Spain\\
$^{3}$Departamento de Astrof\'{i}sica, Universidad de La Laguna, E-38206 La Laguna, Tenerife, Spain\\
$^{4}$National Solar Observatory, 3665 Discovery Drive, Boulder, CO 80303, USA\\
$^{5}$Max-Planck-Institut f\"ur Sonnensystemforschung, Justus-von-Liebig-Weg 3, 37077 G\"ottingen, Germany\\
$^{6}$School of Space Research, Kyung Hee University, Yongin, Gyeonggi, 446-701, Republic of Korea\\
$^{7}$Kiepenheuer-Institut f\"ur Sonnenphysik, Sch\"oneckstr. 6, 79104 Freiburg, Germany
}
\email{iker@iac.es}


%\altaffiltext{1}{Instituto de Astrof\'{i}sica de Canarias, Avda. V\'{i}a L\'{a}ctea s/n, E-38205 La Laguna, Tenerife, Spain}
%\altaffiltext{2}{iker@iac.es}

\begin{abstract}

We study the relation between mesogranular flows, convectively driven sinks and magnetic fields using high spatial resolution spectropolarimetric data acquired with the Imaging Magnetograph eXperiment on board \textsc{Sunrise}. We obtain the horizontal velocity flow fields of two quiet-Sun regions (31.2\,$\times$\,31.2\,Mm$^{2}$) via local correlation tracking. Mesogranular lanes and the central position of sinks are identified using Lagrange tracers. We find $6.7\times10^{-2}$\,sinks per Mm$^{2}$ in the two observed regions. The sinks are located at the mesogranular vertices and turn out to be associated with (1) horizontal velocity flows converging to a central point and (2) long-lived downdrafts. The spatial distribution of magnetic fields in the quiet Sun is also examined. The strongest magnetic fields are preferentially located at sinks. We find that 40\,\% of the pixels with longitudinal component of the magnetic field stronger than 500\,G are located in the close neighborhood of sinks. In contrast, the small-scale magnetic loops detected by Mart\'{i}nez Gonz\'{a}lez et al. in the same two observed areas do not show any preferential distribution at mesogranular scales. The study of individual examples reveals that sinks can play an important role in the evolution of quiet-Sun magnetic features. 

\end{abstract}




\section{Introduction}

Most quiet-Sun magnetic fields evolve on the solar surface driven by convective motions. The largest magnetic structures outline the boundaries of supergranular cells ---the magnetic \textit{network}. Inside the supergranular cells smaller magnetic flux concentrations of both polarities permeate the solar \textit{internetwork}. They tend to concentrate in mesogranular lanes \citep{2011ApJ...727L..30Y} and a significant fraction of the magnetic flux emerges into the surface co-spatially with granules in the form of small-scale magnetic loops \citep{2007A&A...469L..39M,2007ApJ...666L.137C,2009ApJ...700.1391M,2010ApJ...714L..94M,2010ApJ...723L.149D,2012ApJ...755..175M}. 

Convection displays highly localized downdrafts where cold plasma returns to the solar interior \citep{1990ARA&A..28..263S,1998ApJ...499..914S,2009LRSP....6....2N}. Due to conservation of the angular momentum, a vortex can be formed as the plasma approaches the downdraft \citep[the \textit{bathtub} effect,][]{1985SoPh..100..209N}. In fact, small-scale vortex flows are ubiquitous in simulations of solar surface convection \citep[e.g.,][]{2011ApJ...727L..50K,2011A&A...526A...5S,2011A&A...533A.126M,2012ASPC..456....3S}. The plasma can also drag magnetic fields toward the draining point, where they can be intensified up to kG values \citep{2010A&A...509A..76D,2010ApJ...719..307K}. As a consequence, a large amount of vorticity can be generated through the interaction of plasma and magnetic fields in the intergranular juctions \citep{2011A&A...526A...5S}. Vertical vorticity is known to concentrate preferentially in negative divergence areas, i.e., downflow regions \citep{1995ApJ...447..419W,2005HvaOB..29...61P,2007CEAB...31...11P}. Vortex flows are observed at large scales \citep[up to 20\,Mm,][]{1988Natur.335..238B,2009A&A...493L..13A} in supergranular junctions and at smaller scales \citep[$\lesssim$\,0.5\,Mm,][]{2008ApJ...687L.131B,2010ApJ...723L.139B,2011MNRAS.416..148V} in granular ones. Small-scale whirlpools are also visible in the chromosphere \citep{2009A&A...507L...9W,2016A&A...586A..25P} and their imprints have been identified in the transition region and low corona \citep{2012Natur.486..505W}. 
 
Vortex flows harboring magnetic fields are rather abundant \citep{2010ApJ...723L.139B}. In particular, \citet{2010A&A...513L...6B} showed evidence of small-scale magnetic concentrations being dragged toward the center of a convergence flow point. The same event was further studied by \citet{2015SoPh..290..301V}, who found sudden downflows and intensification processes of the magnetic concentrations. Magnetic field enhancements preceded by strong downdrafts are generally understood in the context of \textit{convective collapse} \citep{1978ApJ...221..368P,1978SoPh...59..249W,1979SoPh...61..363S,1979SoPh...62...15S}. \citet{2011A&A...531L...9M} found that the trajectories of some small-scale loop footpoints describe a vortical motion suggesting that they may be engulfed by a downdraft. Recently, \citet{2014ApJ...789....6R} observed a loop footpoint being advected and concentrated in a point-like sink together with other same-polarity weak magnetic patches. All these papers, and the previous knowledge that intergranular magnetic fields tend to concentrate near the intersections of multiple granules, strengthen the idea that localized downdrafts\footnote{We shall use the term sink throughout the paper as a synonym of these localized downdrafts.} are places where the concentration of magnetic fields is favored. In order to confirm such a relation, quantitative and statistical information is still required.

The aim of this work is to provide a quantitative basis for the association between convectively driven point-like sinks and small-scale magnetic fields in the quiet Sun. We use spectropolarimetric observations from the Imaging Magnetograph eXperiment \citep[IMaX;][]{2011SoPh..268...57M}  on board the \textsc{Sunrise} balloon-borne solar observatory \citep{2010ApJ...723L.127S,2011SoPh..268....1B,2011SoPh..268..103B,2011SoPh..268...35G}. IMaX provides stable time series of both intensity and polarization filtergrams at high spatial resolution ($\sim$\,100\,km). Using the same \textsc{Sunrise}/IMaX data, \citet{2011ApJ...727L..30Y} clearly demonstrated that magnetic elements are preferentially located in mesogranular lanes and \citet{2010ApJ...723L.139B} observed a large number of magnetic features swirling in convectively driven vortex flows. In the latter work, they detected vortices from the horizontal velocities obtained through local correlation tracking (LCT) of magnetograms and other IMaX images. They found a mean vortex duration of about 8\,minutes and several events appearing at the same location in the course of the time series.  The recurrent events were also observed in the flow field maps when averaged over the entire data set. In the present paper we find that such long-lived sinks are located in mesogranular junctions and that they are places where the strongest magnetic fields tend to concentrate.

\section{Observations}

\begin{figure*}
\includegraphics[width=\textwidth]{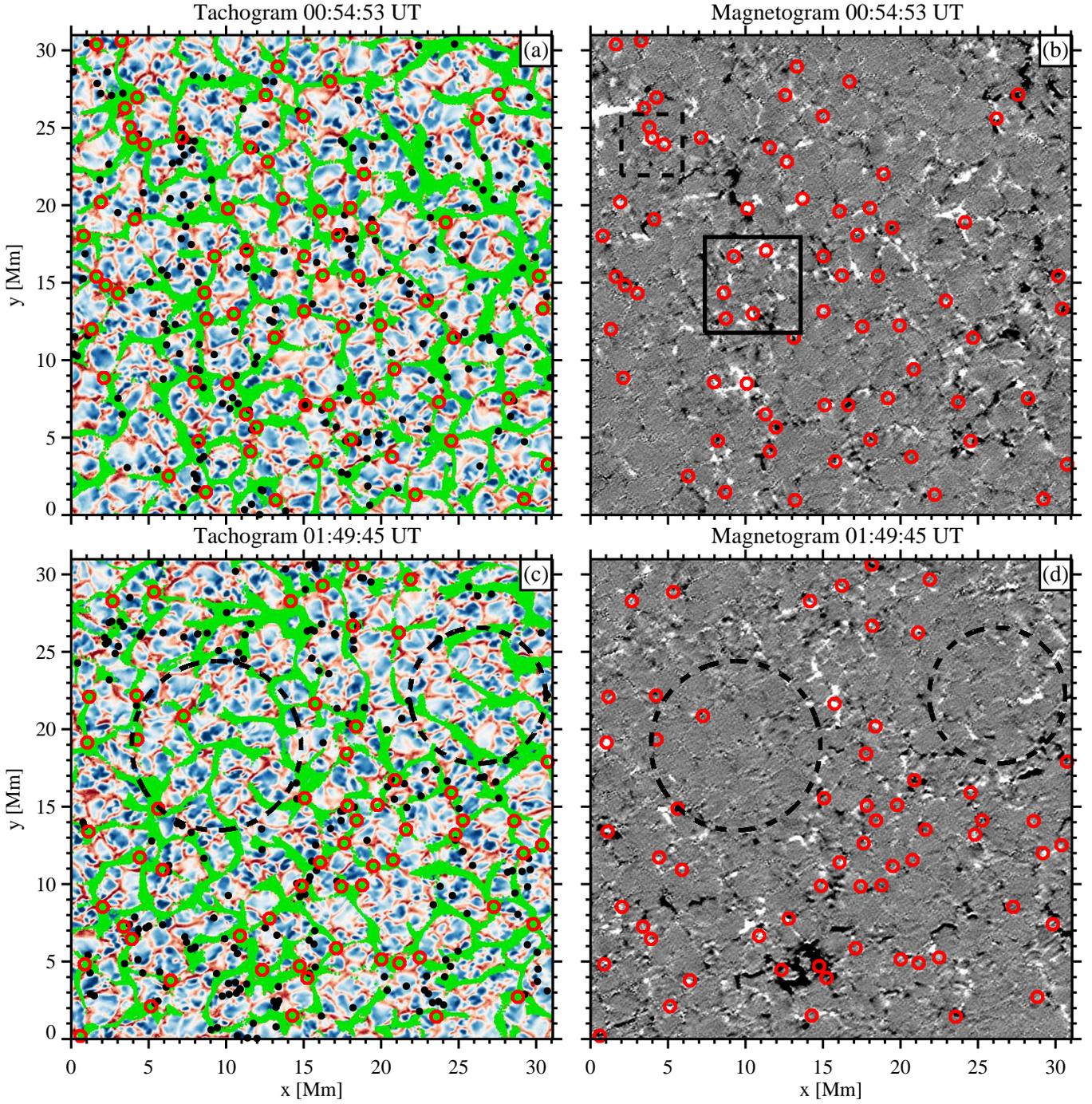}
\caption{Left panels: green pixels represent locations with more than 2 corks pixel$^{-1}$ at $t=21.6$\,minutes and its 1 pixel neighborhoods for time series S$_{1}$ (a) and S$_{2}$ (c), respectively. The background image is the LOS velocity at $t=18.3$\,minutes (saturated at $\pm$\,1.5\,km\,s$^{-1}$), where red and blue regions correspond to downflow and upflow areas, respectively. Black filled circles show the average position (between the two footpoints) of the small-scale loops detected by \citet{2012ApJ...755..175M}. Right panels: the longitudinal component of the magnetic field, $B\,\cos\,\gamma$, at $t=18.3$\,minutes (saturated at $\pm$\,50\,G) for S$_{1}$ (b) and S$_{2}$ (d), respectively. Red circles (with a radius of 9 pixel $\sim$\,360 km) represent the positions of the 131 sinks. The two areas delineated by the dashed circles mark two \textit{dead calm} regions found by \citet{2012ApJ...755..175M}. The black dashed box represents the region analyzed in \citet{2014ApJ...789....6R}. The black solid rectangle, with a FOV of 6.2\,$\times$\,6.2\,Mm$^{2}$, illustrates the area used in Figure \ref{fig4}.}
\label{fig1}
\end{figure*}

We use high-quality spectropolarimetric data obtained with IMaX on board the \textsc{Sunrise} balloon-borne solar observatory. IMaX is a dual-beam imaging spectropolarimeter with full Stokes vector capabilities in the Fe\,\textsc{i} line at 525.02\,nm  (Land\'{e} factor g\,=\,3). The line is sampled by a Fabry--P\'{e}rot interferometer at five wavelength positions located at $\lambda$\,=\,$-$8,\,$-$4,\,$+$4,\,$+$8, and $+$22.7\,pm from the line center \citep[V5-6 mode of IMaX; see][for details]{2011SoPh..268...57M}. The polarization analysis is performed by two liquid crystal variable retarders and a beam splitter. The image sequences were recorded close to a quiet area at disk center on 2009,  June 9. We analyze two different time series, namely, \textit{S$_{1}$}, from 00:36:03 to 00:58:46\,UT (22.7\,minutes duration) and \textit{S$_{2}$}, from 01:30:54 to 02:02:29 UT (32.1\,minutes), with a cadence of 33.25\,s, and a spatial sampling of 39.9\,km. 

The science images were reconstructed using phase diversity measurements \citep{1982OptEn..21..829G,1996ApJ...466.1087P} as described by \citet{2011SoPh..268...57M}. This procedure effectively reduces the IMaX field of view (FOV) down to about 43\arcsec\,$\times$\,43\arcsec (31.2\,$\times$\,31.2\,Mm$^{2}$). After reconstruction, the spatial resolution has been estimated to be 0\farcs 15--0\farcs 18 and the noise level in each Stokes parameter is about 3\,$\times$\,10$^{-3}\,I_{\rm c}$ ($I_{\rm c}$ being the continuum intensity). We recover information of the vector magnetic field and line-of-sight (LOS) velocities through inversions of the full Stokes vector using the SIR code \citep{1992ApJ...398..375R} as described in \citet{2015ApJ...810....79R}. Height independent values for the three components of the magnetic field and LOS velocity are assumed. The magnetic filling factor (i.e., the fraction of
the pixel filled with magnetic field) is set to unity. From the magnetic field strength $B$ and the inclination $\gamma$ we derive the longitudinal component of the magnetic field (hereafter referred to as the longitudinal magnetogram) $B_{\rm long}\,=\,B\,\cos\,\gamma$.

\section{Convectively driven sinks}
\label{sec3}

\subsection{Identification of mesolanes}
\label{sec31}

Mesogranulation is a horizontal cellular flow pattern revealed through the LCT technique when applied to intensity images of the solar granulation \citep{1981ApJ...245L.123N,1988ApJ...327..964S,1989ApJ...336..475T,1991A&A...241..219B,1992Natur.356..322M,1998A&A...330.1136R,2011ApJ...727L..30Y}. Here we search for this pattern in the continuum intensity filtergrams of the two data sets. We use a common time coverage for both time series, namely, the total duration of series S$_{1}$ (the shorter one), so that we analyze only the first 42 snapshots of S$_{2}$. This criterion will allow us to equally define inter-mesogranular lanes (called here mesolanes for short) in both time series. We apply a p-mode subsonic filter \citep{1989ApJ...336..475T} to remove the 5-min solar oscillations. This process degrades the first and the last frames, which are removed from our time series. The final data sets last, therefore, 21.6\,minutes. We employ the LCT technique \citep{1988ApJ...333..427N} as implemented by \citet{1994IRN..31} to obtain the mean horizontal velocity field averaged over the duration of the data sets. This technique correlates small local windows in consecutive images to find the best-match displacement. The tracking window is defined by a Gaussian function with a FWHM\,=\,600\,km. After measuring the horizontal velocity vector $\textbf{\textit{v}}=\textbf{\textit{v}}_{x}+\textbf{\textit{v}}_{y}$, we also compute the flow divergence $\nabla\,\textbf{\textit{v}}=\frac{\partial \textbf{\textit{v}}_{x}}{\partial x}+\frac{\partial\,\textbf{\textit{v}}_{y}}{\partial y}$ and vertical vorticity $(\nabla\,\times\,\textbf{\textit{v}})_{z}=\frac{\partial \textbf{\textit{v}}_{y}}{\partial x}-\frac{\partial \textbf{\textit{v}}_{x}}{\partial y}$. 

We outline the location of mesolanes by using Lagrange tracers (\textit{corks}). Employing the mean horizontal velocity field, we track the evolution of each individual cork as described in \citet{1992ApJ...727L..30Y}. Initially, a cork is located at each pixel of the image. Then the nearest neighbor velocity vector is used to ``move'' the corks. The trajectories are integrated in time assuming a time step of 33.25\,s. For convenience, this value has been chosen to be the same as the cadence of the time series. With this choice the mesogranular pattern is clearly visible after 21.6\,minutes (i.e., 39 time steps), when most of the corks have already converged to mesolanes. The thickness and number of mesolanes depends significantly on the time the corks are allowed to move. After 21.6\,minutes we find a good compromise between the smallest thickness and maximum number of mesolanes. At that time step we count the number of corks in each pixel to obtain the cork density function, $\rho_{cork}$ \citep{2011ApJ...727L..30Y}, and we define the mesolanes as the locations where $\rho_{cork}\,\geq\,2$, augmented by a 1-pixel wide neighborhood. These areas are marked in Figure \ref{fig1} (a) and (c) through green pixels overlaid over the LOS velocity maps (tachograms) at $t=18.3$\,minutes for time series S$_{1}$ and S$_{2}$, respectively. They clearly delineate a fully developed network of mesogranular cells with a size of 5-10\,Mm \citep{1981ApJ...245L.123N}.

\subsection{Identification of converging flows}

The mesogranular pattern is visible after only 21.6\,minutes. However, if the corks are tracked for even longer,  they finally end up (after some 5\,hours\footnote{Notice that the duration of the cork movie is unrelated to the length of the observations. The former has been computed by tracking cork trajectories through the mean horizontal velocity field as described in Section \ref{sec31}.}) in well localized points, in places where the horizontal velocity vectors converge on a central point. We obtain therefore the central positions of persistent \textit{converging flows}, which appear in the mean flow corresponding to the full time series. We identify a total of 131 converging horizontal flows in the two observed regions. Taking into account the spatial area covered by each FOV, we get a density of $6.7\,\times\,10^{-2}$\,converging\,flows per Mm$^{2}$.

\subsection{Classification of converging flows}

In previous studies, small whirlpools were first detected as swirling motions of bright points \citep{2008ApJ...687L.131B}. Later they were identified from LCT horizontal velocities of magnetograms \citep{2010ApJ...723L.139B} and G-band filtergrams \citep{2011MNRAS.416..148V}. In the same way as done here, \citet{2011MNRAS.416..148V} applied LCT to the whole FOV and duration of two G-band time series (20\,minutes each) acquired with the Swedish 1-m solar telescope. By visual inspection of the horizontal flow-field maps, they identified individual vortices as those locations where the horizontal velocity vectors converge to a central point and form a swirl. They found these regions to be coincident with the final destination of corks. However they detected many other places where the corks got accumulated without any apparent swirling motion.

\begin{figure}[!t]
\includegraphics[scale=0.66]{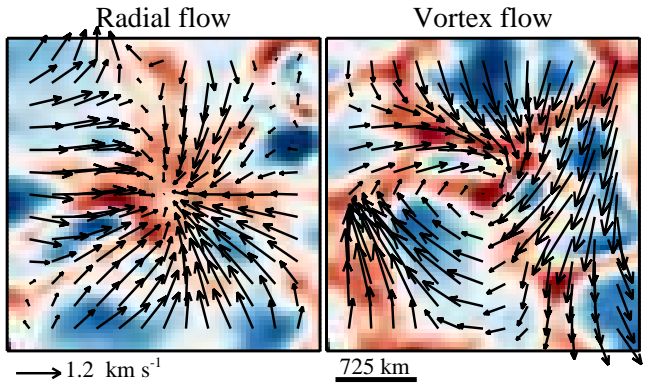}
\caption{Two examples of converging flows detected in the IMaX data. The background images represent the LOS velocity at $t=18.3$\,minutes (saturated at $\pm$\,1.8\,km\,s$^{-1}$). Red and blue regions corresponds to downflow and upflow areas, respectively. The black arrows represent the horizontal velocity vectors obtained from LCT. The left  panel shows an example of what we name a \textit{radial flow} where velocity vectors point radially towards the center of the image. The right panel displays a \textit{vortex flow} where the velocity vectors exhibit a swirling motion with a clockwise sense of rotation.}
\label{fig2}
\end{figure}

Here we inspect both the horizontal flow maps and the cork movies to analyze all the detected convergence centers. We realize that the corks follow different trajectories on their way to being engulfed by the converging flows. Some converge radially while other trace a spiral path as they fall into the inflow centers. In Figure \ref{fig2} black arrows display the distribution of horizontal velocities for examples of  two types of converging flows. In the left panel, the vectors point radially towards the center of the image, while in the right panel, they display a swirl. We shall hereafter refer to them as \textit{radial flows} and \textit{vortex flows}, respectively. We classify all the converging flows into these two categories by visual inspection. They are labeled as vortex flows when a sense of rotation can be ascribed to them from both cork movies and horizontal flow maps. In all other cases, they are designated as radial flows.

\subsection{Statistics of converging flows}
\label{sec34}

We detect 46 vortex flows (35\,\% of the total sample of converging flows) and 85 radial flows. These result in densities of $2.4\,\times\,10^{-2}$\,vortex flows per Mm$^{2}$ and $4.4\,\times\,10^{-2}$\,radial flows per Mm$^{2}$, respectively. Our values for vortices is comparable to the number obtained by \citet{2011MNRAS.416..148V} of 2.8--3.1\,$\times 10^{-2}$\,vortices per Mm$^{2}$. 

\begin{figure}[!t]
\includegraphics[scale=0.66]{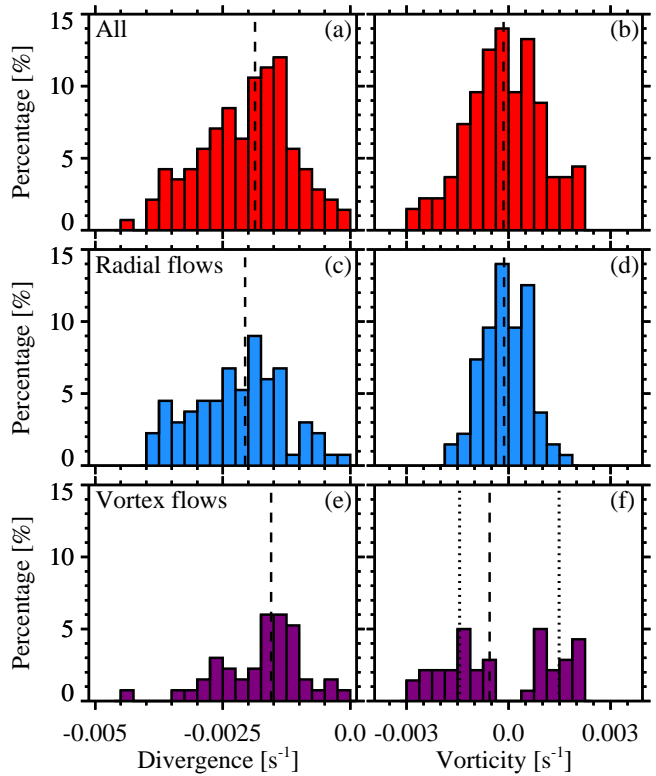}
\caption{Histograms of divergence (a, c, and e) and vertical vorticity (b, d, and f) obtained from LCT proper motions at the convergence centers. Top panels are for all the 131 converging flows, middle panels are for the 85 radial flows, and bottom panels are for the 46 vortex flows. The vertical dashed lines indicate the median values of the distributions. In panel (f) the left and right vertical dotted lines mark the median value of the clockwise and counterclockwise vortices, respectively. We use a bin size of  $2.5\times 10^{-4}$\,s$^{-1}$ and $3.75\times 10^{-4}$\,s$^{-1}$ for divergence and vertical vorticity, respectively.}
\label{fig3}
\end{figure}

\begin{figure}
\includegraphics[scale=0.95]{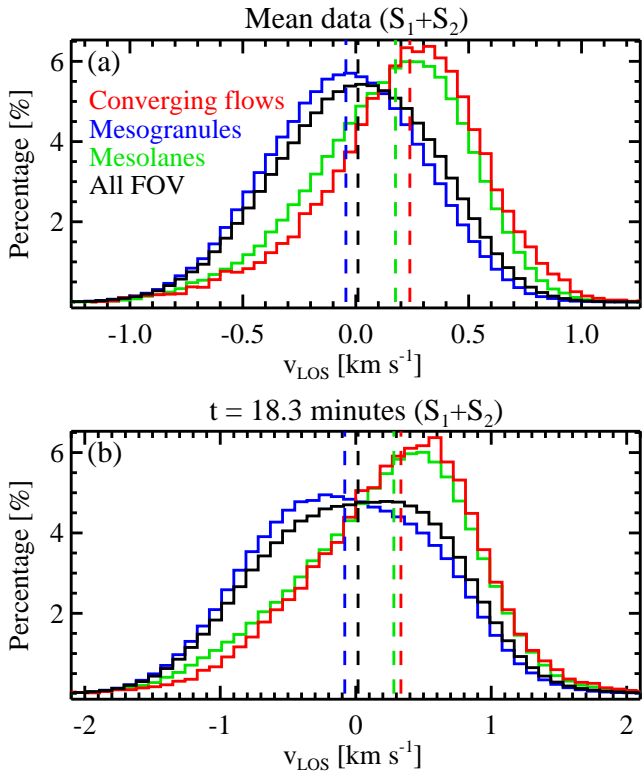}
\caption{Histograms of the LOS velocity for combined values within both observed areas (S$_{1}+$S$_{2}$). Panel (a) plots mean values over the time series, whereas panel (b) shows values within a single snapshots at $t=18.3$\,minutes after the start of the time series. Solid lines stand for all pixels in the FOV (black), converging flows (red), mesolanes (green), and mesogranules (blue), respectively. These regions are defined in the text. The corresponding vertical dashed lines indicate the median values of the distributions.}
\label{fig4}
\end{figure}

\begin{figure}[!t]
\includegraphics[scale=0.95]{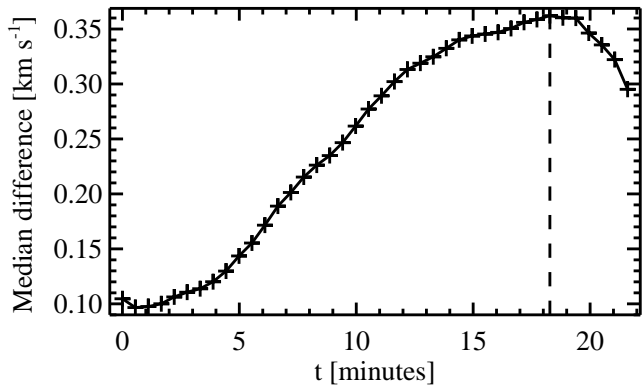}
\caption{Differences between the LOS velocity median values at mesolanes and mesogranules over the duration of the time series. The vertical dashed line indicates the moment ($t=18.3$\,minutes) where the difference is largest.}
\label{fig5}
\end{figure}

In Figure \ref{fig3} we display histograms of the divergence (left panels) and the vertical vorticity (right panels) at the convergence centers. By construction, all the converging flows (panel a) are located in negative divergence areas with a median value of $-1.9\times 10^{-3}$\,s$^{-1}$. The divergence distribution is very similar for both vortex flows (panel e) and radial flows (panel c) with a 25\,\% smaller median value for the former, suggesting that vortex sinks are (25\,\%) less vigorous than uniform ones. Not surprisingly, the distribution of the vertical vorticity is completely different. Panel (d) shows that most of the radial flows have very low vorticity, whereas the negative and positive humps in panel (f) are produced by the distribution of whirls with clockwise and counterclockwise sense of rotation, respectively.\footnote{Since our classification is visual there is no clear threshold in vertical vorticity that separates vortex flows from radial flows.} Both clockwise and counterclockwise vortices have a median absolute vertical vorticity of $1.5\times 10^{-3}$\,s$^{-1}$. We find a slightly larger number of clockwise (54\,\%) than counterclockwise motions (46\,\%), but the difference is not significant in view of the number of studied events.

\subsection{LOS velocities in converging flows}

If the computed horizontal flows are really describing true mass motions, then the horizontal proper motions should be related to vertical velocities. In particular, positive and negative divergences (i.e., mesogranules and mesolanes) must be correlated with upflows and downflows, respectively. Equivalently, converging flows should be coincident with downdrafts if they are indeed related to localized sinks.

In previous works \citep{2008ApJ...687L.131B,2010ApJ...723L.139B,2011MNRAS.416..148V}, the authors assumed without further examination that converging flows are locations of sinking gas. Here, we use the capabilities of IMaX to determine the LOS velocities at these locations. For this purpose, we build a binary mask by assuming that the spatial size of a converging flow is given by a 6\,pixel ($\sim$\,240\,km) radius circle plus a 3\,pixel wide annulus neighborhood. This 240\,km radius corresponds to the mean radius found for converging flows by \citet{2011MNRAS.416..148V}. In addition, we divide the FOV into mesogranules and mesolanes. The former are defined as those areas in the FOV that are not covered by the latter (defined in Section \ref{sec31}). We then compare the histograms of the LOS velocities inferred from the SIR inversions at mesogranules, mesolanes and converging flows with that for the whole FOV. 

The position of mesolanes and converging flows depend on the LCT average and therefore their comparison with individual tachograms of the time series is not straightforward. In a first instance, we compare them with the mean LOS velocities (averaged over the duration of the time series) at both observed regions (S$_{1}+$S$_{2}$). The resulting histograms are shown in Figure \ref{fig4} (a). Different line colors indicate different locations, namely, red for converging flows, blue for mesogranules, green for mesolanes, and black for the whole FOV. The corresponding vertical dashed lines indicate the median values of each distribution. Mesogranules are blueshifted and mesolanes redshifted with median  values of -0.05 and 0.17\,km\,s$^{-1}$, respectively. The redshift at convergence areas (red line) is even larger than in mesolanes, with a median value of 0.24\,km\,s$^{-1}$. On average, mesogranules appear associated with upflows while mesolanes and converging flows  are mostly located at downflows.

If we do the same for individual snapshots of the time series, we find that the correlation between mesolanes and downflows is highest at $t=18.3$\,minutes after the start of the time series. This is shown in Figure \ref{fig5} where we display the differences between the median LOS velocity values at mesolanes and mesogranules over the time series. The vertical dashed line indicates the instant where the difference is largest.  The corresponding histograms at $t=18.3$\,minutes are shown in Figure \ref{fig4} (b). The results are similar to those obtained in Figure \ref{fig4} (a). In this case, mesogranules (blue line), mesolanes (green line) and converging flows (red line) are characterized by median velocities of -0.1, 0.27, and 0.33\,km\,s$^{-1}$, respectively. Figure  \ref{fig2} shows two representative examples of converging flows associated with downflows. 

In summary, we can say that mesolanes and in particular converging flows are (1) preferentially located within intergranular lanes and (2) associated with long-lived  downdrafts. Therefore, we have shown that converging flows are indeed locations of sinking gas and we shall hereafter refer to them as \textit{convectively driven sinks}.

At this point, we examine the question of whether mass is conserved at sinks. For that to occur, the lateral influx of mass over a scale height $H$ should be equal to the vertical mass flux. Approximating the sink geometry by that of a circular cylinder of radius $R$ and height $H$, mass conservation thus requires
\begin{equation}
2\pi RH\rho v_R\approx\pi R^{2}\rho v_z.
\end{equation}
To evaluate the lateral and vertical mass fluxes we need the radial velocity at the edge of the sink, $v_R$, and the vertical velocity, $v_z$. Assuming rotational invariance, the flow divergence can be expressed as $\nabla\,\textbf{\textit{v}}=\frac{1}{r}\,\frac{d(r\,\textbf{\textit{v}}_r)}{dr}$ in cylindrical coordinates. Taking $v_r=\frac{r}{R}\,v_R$ yields $v_R=\frac{R\,\nabla\,\textbf{\textit{v}}}{2}$. With $R=$\,240\,km and $\nabla\,\textbf{\textit{v}}=1.9\times\,10^{-3}$\,s$^{-1}$ (see Section \ref{sec34}), we obtain $v_R=0.23$\,km\,s$^{-1}$. The influx of mass is then $2\pi RH\rho v_R=5.2\times10^{19}\,\rho$\,[g\,s$^{-1}$] over a density scale height of 150\,km \citep[e.g.,][]{1971SoPh...18..347G}. On the other hand, taking $v_z=0.33$\,km\,s$^{-1}$ yields a vertical mass flux $\pi R^{2}\rho v_z=6\times 10^{19}\,\rho$\,[g\,s$^{-1}$], which is in good agreement with the lateral mass influx. Thus, we conclude that mass is approximately conserved in sinks.

\subsection{Spatial distribution of sinks}

In Figure \ref{fig1} we show the distribution of sinks (red circles) overlaid on the tachograms and longitudinal magnetograms of the two observed areas at $t=18.3$\,minutes. Green pixels in the tachograms represent the distribution of corks delineating the mesolanes. The sinks show a mesogranular distribution as most of them are located at the junctions of multiple mesogranular lanes. However, their spatial distribution is not completely uniform and there are some extended areas lacking detected sinks.
There are two particularly prominent regions in time series S$_{2}$. Interestingly they are coincident with the \textit{dead calm} areas found by \citet{2012ApJ...755..175M}, characterized by very low magnetic flux and a lack of small-scale magnetic loops. We have marked them with dashed circles in Figures \ref{fig1} (c) and (d). The two regions contain several mesogranules, but only a single sink is observed in the largest void. Visual inspection of the sinks in the longitudinal magnetograms (Figures \ref{fig1} b and d) already reveals that many sinks harbor magnetic fields. This connection between magnetic fields and sinks gets even more evident if one looks at the evolution of magnetic elements. As we will describe in the following Section, magnetic structures are seen to be affected by convectively driven sinks.

\section{The evolution of magnetic features driven by convective motions}
\label{sec4}

\begin{figure*}[!ht]
\includegraphics[width=\textwidth]{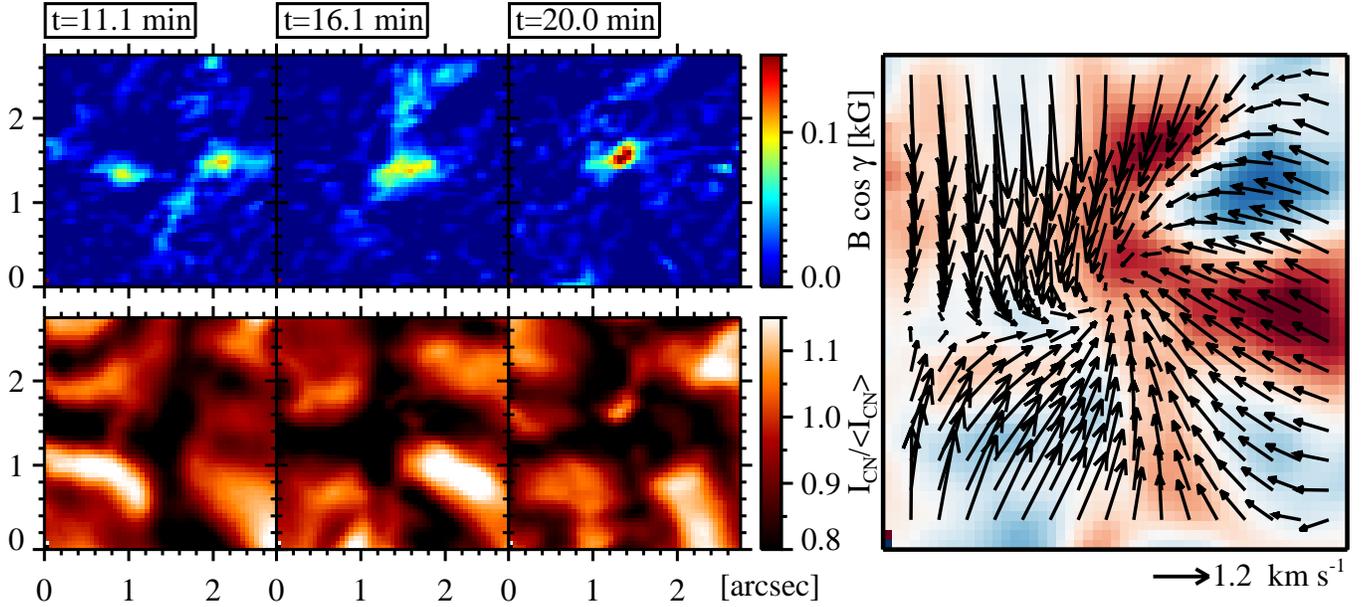}
\caption{Temporal sequence of a coalescence process. Upper row: IMaX Fe\,\textsc{i}\,525.02\,nm longitudinal magnetic field. Lower row: SuFI images in the CN bandhead, centered at 388\,nm with FWHM\,$\approx$\,0.8\,nm. The rightmost panel shows the horizontal velocity maps (black arrows) derived through the LCT technique averaged over the whole time series of 21.6\,minutes.  The background image shows the mean tachogram  averaged over the whole time series and saturated at $\pm$\,0.8\,km\,s$^{-1}$. The black arrows display the horizontal velocity vectors. The length of the black arrow below the lower right corner of the figure corresponds to 1.2\,km\,s$^{-1}$.}
\label{fig6}
\end{figure*}

This Section presents examples of three distinct processes that affect the evolution of quiet-Sun magnetic features: coalescence, cancellation and fragmentation. The statistics  of such processes in the \textsc{Sunrise}/IMaX data set has been studied in detail by \citet{2016arXiv160808499A}. Here we show that these evolutionary processes are driven by convective motions. In particular, coalescence and cancellation of magnetic structures are commonly observed at the sinks. Generally, fragmentation also takes place at sinks. At these locations, however, the resulting fragments are not usually completely detached from each other. Rather, the converging flows tends to merge them again into a single magnetic element. This phenomenon has already been  observed by \citet{2015ApJ...810....79R}: multi-cored magnetic structures are seen to live as a single entity for long periods; the individual cores merge and split. Our examples of fragmentation (see Section \ref {sec43}) are taken from such multi-cored structures. Outside sinks, the complete splitting of magnetic elements is widely observed when the features are ``squeezed'' between two converging granules. Note that only illustrative examples are presented and discussed here, whose relevance for processes taking place in other magnetic features will need to be investigated on the basis of a much larger sample. 

\subsection{Coalescence} 

Figure \ref{fig6} shows a coalescence process with a selection of three longitudinal magnetograms (upper row) and co-aligned CN-band filtergrams (bottom row). The CN-band filtergrams were acquired with the \textsc{Sunrise} Filter Imager \citep[SuFI;][]{2011SoPh..268...35G} and have been properly aligned with IMaX images as described in \citet{2015ApJ...810....79R}.  The rightmost panel displays the horizontal velocity vectors obtained through the LCT technique averaged over the whole time series of 21.6\,minutes (see Section \ref{sec3}). The background image shows the mean tachogram averaged over the same time. 

The velocity map reveals a sink as the flows converge towards the center of the image where a downflow is present. In the first magnetogram two weak same-polarity magnetic patches are seen. As time goes by the patches move towards the convergence center where they merge into a single more intense magnetic element. During this coalescence process the longitudinal magnetic field increases from 100\,G to about 200\,G and a bright point (BP) is also formed as seen in the CN-band filtergrams. Contrary to the standard flux-tube theory \citep{1976SoPh...50..269S}, we find CN-band BPs related to magnetic elements with fields weaker than 1\,kG (see also Section \ref{sec43}). Based on a  comparison between the same \textsc{Sunrise} observations and MHD simulations, \citet{2014A&A...568A..13R} concluded that all magnetic BPs are indeed associated with kG fields provided they are spatially resolved at the resolution of \textsc{Sunrise}/IMaX. However, from a simultaneous inversion of near infrared and visible spectral lines with a free magnetic filling factor, \citet{2007A&A...472..607B} found that the field strength of G-band BPs has a flat distribution from 500\,G to 1500\,G.

\begin{figure*}
\includegraphics[width=\textwidth]{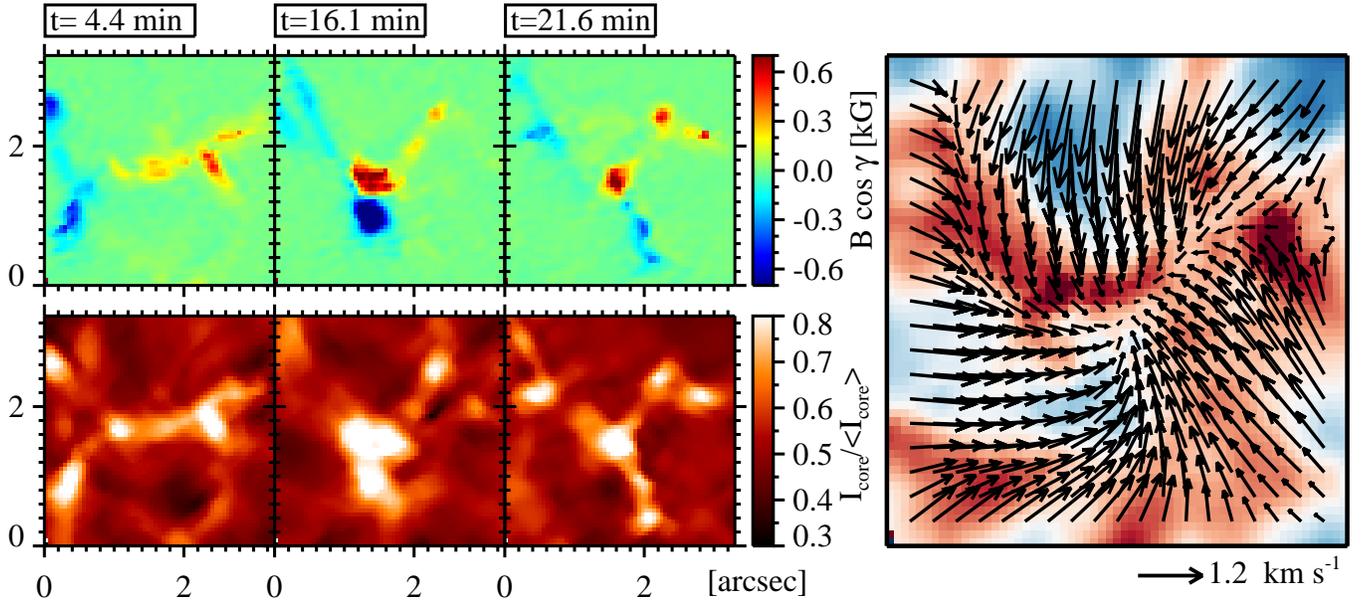}
\caption{Same as Figure \ref{fig6}, during a partial cancellation process. Here the lower row shows the line-core intensity of the IMaX Fe\,\textsc{i} line at 525.02\,nm in units of the continuum intensity. Please also note the different color scale in the upper row of images.}
\label{fig7}
\end{figure*}

\begin{figure*}
\includegraphics[width=\textwidth]{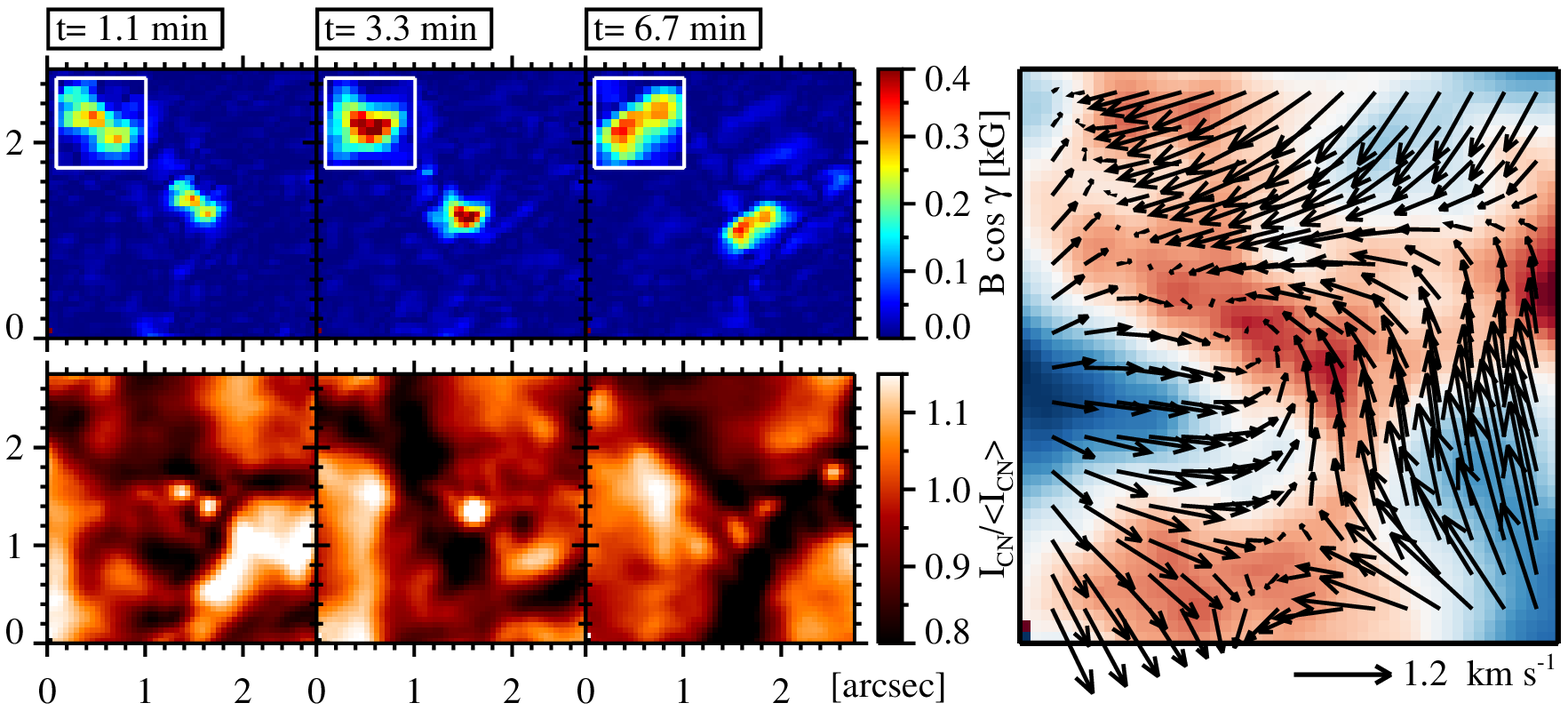}
\caption{Same as Figure \ref{fig6}, during the evolution of a  multi-cored magnetic structure. White boxes in the upper row display a zoom of the magnetic structure.}
\label{fig8}
\end{figure*}

\subsection{Cancellation} 

Figure \ref{fig7} shows a partial cancellation process. The bottom row displays the line-core intensity instead of a CN-band filtergram as the event is located outside the FOV of SuFI. The line-core intensity has been determined by fitting a Gaussian to the observed IMaX Stokes $I$ profiles \citep[see][]{2015ApJ...810....79R}. In the first frame, two diffuse structures of opposite polarities are observed. These structures are concentrated by the horizontal converging flows shown in the velocity map, thus forming two intense magnetic elements that interact with each other (see second frame). At this point, the positive and negative patches have magnetic fluxes of $+\,6.5 \times 10^{17}$\,Mx and $-\,9 \times 10^{17}$\,Mx, respectively. The positive polarity feature survives this interaction intact, although with a greatly reduced magnetic flux of $+\,3.5 \times 10^{17}$\,Mx, while the negative one gets more diffuse and decreases its magnetic flux down to $-\,2.5 \times 10^{17}$\,Mx.

\subsection{Fragmentation and merging in a multi-cored magnetic structure} 
\label{sec43}

Figure \ref{fig8} shows the evolution of a small magnetic feature. It is located within a vortex flow as indicated by the swirl displayed in the horizontal velocity map. In the first image the magnetic feature contains two cores with longitudinal field of 300\,G each. Although these cores are likely spatially unresolved, the co-aligned CN-band image confirms that they are indeed separate as each one appears associated with a CN bright point. This testifies to the outstanding quality of the \textsc{Sunrise}/IMaX spectropolarimetric capabilities. The time sequence shows that the inner cores, and so the related BPs, merge to form a more intense element, which reaches up to 600\,G. The coalescence is probably driven by the sink. Soon afterwards, however, the magnetic feature splits again into two weaker components. The evolution of such multi-cored  magnetic structures has been studied in detail in \citet{2015ApJ...810....79R}. The continuous merging and splitting processes of the inner cores is governed by the evolution of surrounding granules and results in magnetic field oscillations of the global entity.

\begin{figure*}[!t]
\includegraphics[width=\textwidth]{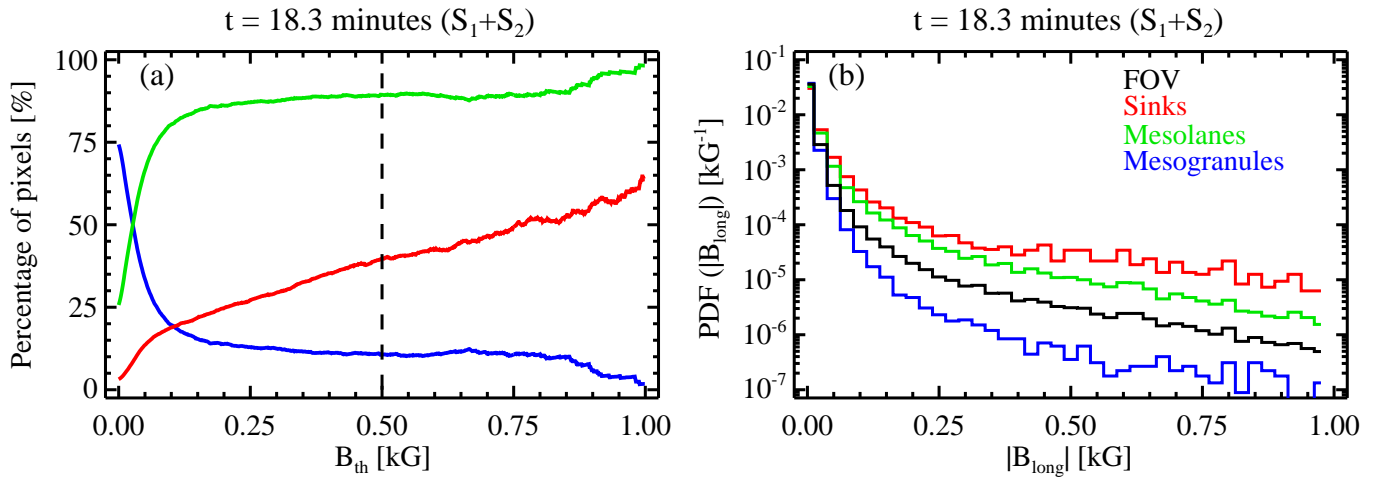}
\caption{Panels (a): percentage of pixels with $\left|B_{\rm long}\right|$ above the value  $B_{\rm th}$ given in the abscissa as obtained in both time series at $t=18.3$\,minutes. Panel (b):  probability density functions of $\left|B_{\rm long}\right|$ at $t=18.3$\,minutes in both time series. The black, red, green, and blue solid lines stand for all pixels in the FOV, sinks, mesolanes, and mesogranules, respectively.}
\label{fig9}
\end{figure*}

\section{Correlation between mesogranules, sinks, and magnetic fields}
\label{sec5}

From visual inspection of the sinks and the study of individual examples \citep[see also][]{2010A&A...513L...6B,2010ApJ...723L.139B,2014ApJ...789....6R,2015SoPh..290..301V}, we obtain indications that strong magnetic features concentrate preferentially at well localized places in the centers of sinks.  In order to obtain a quantitative estimate of the spatial distribution of magnetic fields\footnote{In what follows, we remove from our analysis the prominent positive patch appearing at the top-left edge of the time series S$_1$ (Figure \ref{fig1} (b)). This magnetic structure enters partially into the FOV in the course of the time series, and it is located at the edge of the FOV, where the horizontal flow fields are not properly determined.}, for each threshold value $B_{\rm th}$ in the range $[0, 1000]$\,G, we consider the number of pixels with $\left|B_{\rm long}\right| > B_{\rm th}$ in the FOV and compute the percentage of those pixels that are located within different subareas. The result is displayed in Figure \ref{fig9} (a) for the combination of both observed regions (S$_{1}+$S$_{2}$) at $t=18.3$\,minutes. Different line colors indicate different locations, namely, green for mesolanes, blue for mesogranules, and red for sinks. The curves for mesolanes and mesogranules add up to 100\,\% at each $B_{\rm th}$ value according to our definition (i.e., the sum of both areas cover the entire FOV), whereas the red curve for sinks is just a subregion of mesolanes. It can be seen that the magnetic structures are mainly located at mesolanes. In agreement with \citet{2011ApJ...727L..30Y}, 80\,\% of the pixels harboring fields with longitudinal component larger than 100\,G are located in mesolanes. We also find that the fraction of pixels stronger than $B_{\rm th}$ increases linearly with $B_{\rm th}$ in sinks. Actually, 40\,\% of the pixels with $\left|B_{\rm long}\right| \gtrsim$\,500\,G (see vertical dashed line in Figure \ref{fig9} (a)) are located in the sinks or in their close vicinity. Such a large concentration is very remarkable, since sinks occupy only 3\,\% of the surface area in contrast with the 26\,\%  of the mesolanes. This statement can be better visualized through the corresponding probability density functions (PDFs) displayed in Figure \ref{fig9} (b). All the PDFs peak at 0\,G, and then decrease rapidly toward stronger fields. This decrease is steeper for mesolanes than for sinks, and even more abrupt for mesogranules. All these measurements suggest that the magnetic fields in the quiet-Sun areas are preferentially located at well localized long-lived downdrafts. 

\begin{figure*}
\includegraphics[width=\textwidth]{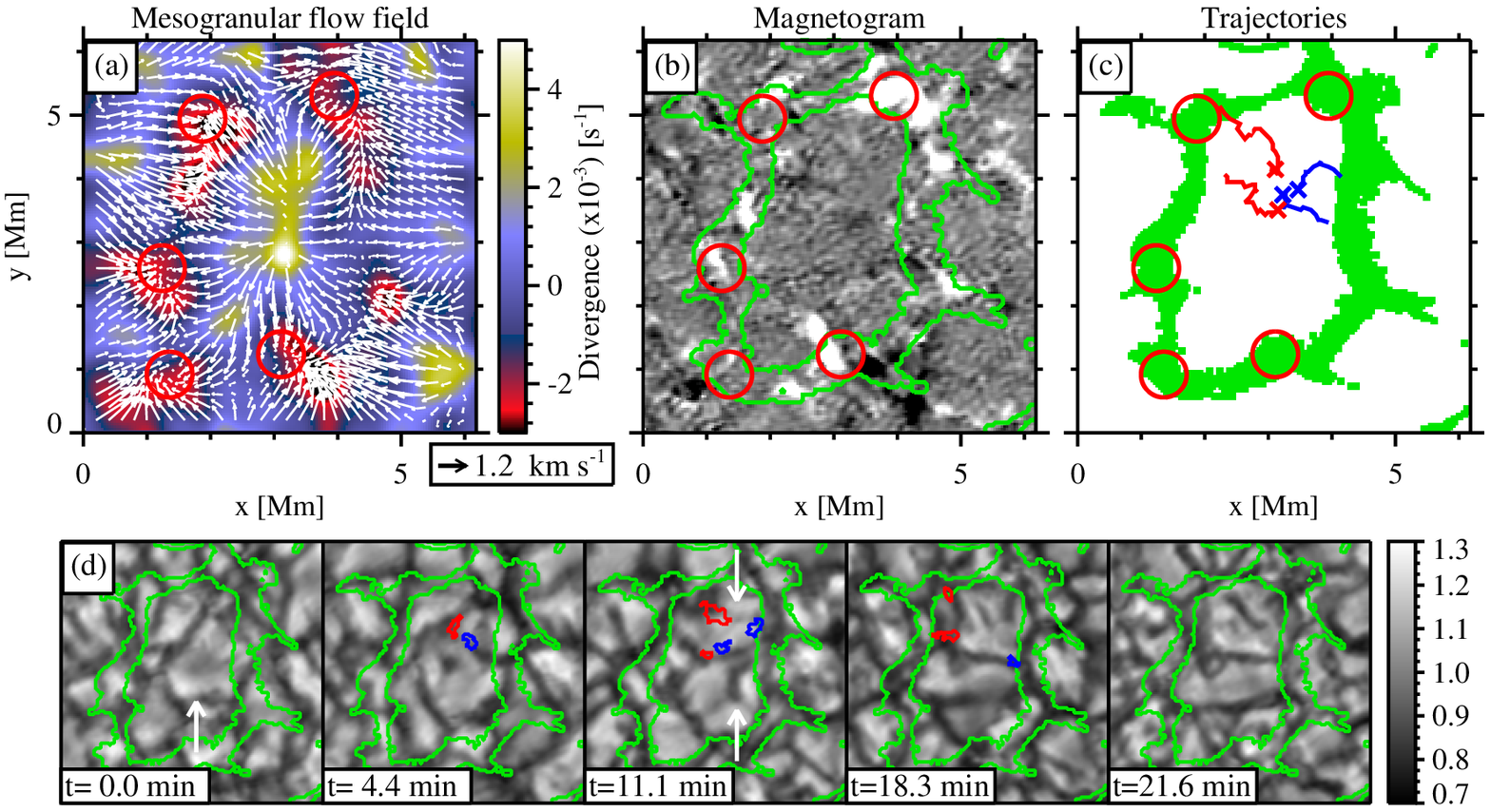}
\caption{Closeup of a mesogranule. The area is located within the black solid rectangle in Figure \ref{fig1}. Panel (a): divergence map (background image) and mesogranular flow field (white arrows). Negative values represent convergence. Panel (b): longitudinal magnetogram at $t=18.3$\,minutes saturated at $\pm$\,50\,G. Green contours delineate the mesolanes. Panel (c): mesolanes (green pixels) and  trajectories followed by the footpoints of initially detected small-scale magnetic loops. Blue and red paths refer to the positive and negative polarity magnetic patches, respectively. Crosses indicate the initial position of the footpoints. Red circles (with a radius of 9 pixel $\sim$\,360\,km) represent the positions of sinks. Panel (d): a time sequence of continuum intensity maps covering the same field of view. Arrows indicate exploding granules. Blue and red contours represent positive and negative polarity magnetic patches, respectively.}
\label{fig10}
\end{figure*}

On granular scales, observations indicate that there is a continuous injection of magnetic flux in the form of small-scale magnetic loops \citep[e.g.,][]{2009ApJ...700.1391M}. Exploring, then, whether any preferential location also applies to these loops is in order. In Figure \ref{fig1} (a) and (c) black filled circles show the distribution of the average position between the footpoints of each detected loop. The locations of loops have been identified by \citet{2012ApJ...755..175M}. The average initial distance between footpoints was found to be $\sim$\,0.25\,Mm. To represent the surface area covered by loops, we have built a binary mask by defining them as 0.25\,Mm diameter circles around the average position between both footpoints. We count the number of pixels in both observed regions and find that 69.5\,\% of them are located in mesogranules, 30.5\,\% in mesolanes, and only 4.5\,\% in sinks. These values are close to those of the surface fraction covered by each region (74\,\% mesogranules, 26\,\% mesolanes, and 3\,\% sinks), suggesting that magnetic loops are homogeneously distributed on mesogranular scales except in the dead calm areas where no loop is seen to emerge \citep{2012ApJ...755..175M}.

\subsection{Mesogranule: a case study}

In Figure \ref{fig10} we zoom in on one of the mesogranules. Panel (a) shows the divergence map with the overlaid white arrows displaying the horizontal flow field. The mesogranule is characterized by a strong positive divergence and a horizontal flow directed from the center of the map outward. The time sequence of the continuum intensity in panel (d) reveals the recurrent appearance of exploding granules. These granules are marked by white arrows. Such families of repeatedly splitting granules can persist for long times \citep[up to 8\,hours,][]{2001SoPh..203..211M,2003A&A...409..299R} and when averaged in time they give rise to positive divergences that are identified as mesogranules \citep{2004A&A...419..757R}.

In panel (b) a longitudinal magnetogram at $t=18.3$\,minutes is shown. The green contours trace the mesolanes and the red circles display the locations of sinks. Notice that in panel (a) the sinks are found at strong negative divergence areas in which the horizontal flows converge.  In the magnetogram, the majority of the magnetic elements are located within the mesolanes and in particular inside or near sinks. As an example, in the bottom right sink two opposite polarity features are found and in the upper right one a prominent magnetic element is observed. The evolution of the latter was studied in detail in \citet{2015ApJ...810....79R}. It contains multiple inner magnetic cores which continuously fragment and merge in response to the evolution of the surrounding granules. Here, we find that this structure is indeed anchored to a persistent downdraft. This constant inward attraction may help maintain all the fragments together in a single magnetic structure. This is just one example of the processes occurring when a magnetic structure is located in a sink (see Section \ref{sec4}). 

In the interior of the mesogranular cell, smaller magnetic patches are also observed. These features are much weaker, with a longitudinal magnetic field component smaller than 50\,G. Some of them have been identified as small loop-like structures because a linear polarization patch appears flanked by two opposite polarities in the longitudinal magnetograms. We have tracked the evolution of their footpoints manually until they merge, cancel with other features, or fade away below the detection limit. We display their trajectories in panel (c). Blue and red lines stand for positive and negative polarity patches and crosses indicate their initial positions. Interestingly, they move radially toward the mesolanes, although they also describe random motions at smaller scales \citep[cf.][]{2011ApJ...743..133A,2011A&A...531L...9M,2014A&A...563A.101J}. In panel (d) we show the evolution of the patches overlaid on continuum intensity maps. Blue and red contours represent positive and negative polarity patches, respectively. They appear located at the edges of granules and are swept to the mesogranular lanes as the granules expand and explode. The uppermost negative polarity patch reaches the periphery of a sink where it soon cancels out with an opposite polarity feature. These two features are observed prior to their cancellation at the upper border of the top left sink in panel (b). 

\section{Discussion and conclusions}

The accuracy and high spatial resolution of the \textsc{Sunrise}/IMaX polarimetric measurements have allowed us to quantify the relation between quiet-Sun magnetic fields, mesogranulation, and convectively driven sinks. We have computed the horizontal flow field through LCT of continuum intensity maps averaged over two time series of roughly 20\,minutes each. From the inferred horizontal velocity vectors, we identified mesogranular lanes and convergence centers by tracking the evolution of Lagrange tracers. Converging flows are preferentially located at the junction of multiple mesograngranular lanes. According to the cork trajectories and the distribution of converging flow fields we  have found two types of \textit{converging flows}, namely, (1) \textit{radial flows}: radially symmetrical flow fields directed inwards to the convergence center; and (2) \textit{vortex flows}: flow fields rotating around a vertical axis. We have detected 131 long-lived converging flows, 65\,\% belonging to type 1, and the rest to type 2. This results in $6.7\,\times\,10^{-2}$\,converging flows per Mm$^{2}$, $4.4\,\times\,10^{-2}$\,radial flows per Mm$^{2}$ and $2.4\,\times\,10^{-2}$\,vortex flows per Mm$^{2}$, respectively. The value for vortex flows is comparable to that found by \citet{2011MNRAS.416..148V}. It is important to remember that our LCT temporal average (21.6\,minutes) and the tracking window (600\,km) is almost identical to those used by \citet{2011MNRAS.416..148V} of 20\,minutes and 725\,km, respectively.

Unlike in previous works, here we have profited from the spectroscopic capabilities of IMaX to determine LOS velocities in converging flows. The LOS velocity histograms reveal that converging flows are (1) preferentially located within intergranular lanes and (2) tend to be  associated with long-lived  downdrafts. Accordingly, converging flows are better described as convectively driven sinks.

By studying individual examples, we have shown that sinks can affect the evolution of magnetic elements. In particular, coalescence, cancellation and fragmentation processes are seen to take place at localized downdrafts. We have also provided quantitative measurements of the relationship between magnetic fields, mesogranules and sinks. First, we have confirmed the finding of \citet{2011ApJ...727L..30Y} that the large majority (80\,\%) of magnetic fields with longitudinal component larger than 100\,G are located in mesogranular lanes. In addition, we have shown that the strongest magnetic features tend to concentrate in long-lived sinks, at the junction of mesogranules. Roughly 40\,\% of the magnetic elements with longitudinal magnetic fields above 500\,G are found within 360\,km of sink centers. In contrast, we found that 69.5\,\% of the 400 small-scale magnetic loops detected by \citet{2012ApJ...755..175M} are located in mesogranules, 30.5\,\% in mesolanes, and only 4.5\,\% in the close neighborhood of long-lived sinks. These values are very similar to the surface fractions covered by each region and we conclude that magnetic loops are homogeneously distributed on mesogranular scales.

We have analyzed the evolution of two particular loops appearing inside a mesogranule. Their footpoints are passively advected by the mesogranular horizontal flows, as they are swept by exploding granules. Through this process they reach the boundaries of the underlying mesogranules and well localized sinks located at the vertices of mesogranular lanes. In such downdrafts they can be confined by converging granular flows and later concentrated up to kG values \citep{2008MNRAS.387..698B,2010A&A...509A..76D,2014ApJ...789....6R}. The advection of weak fields by mesogranular horizontal flows and their concentration in mesogranular vertices is also supported by magnetoconvection models that study the interaction of convective flows with an imposed field \citep[see e.g.,][]{2006ApJ...642.1246S}. However, this scenario is also compatible with a small-scale dynamo \citep{2014ApJ...789..132R}. In fact, small-scale dynamo action is more efficient when mesogranulation is present \citep{2014A&A...562A..72B}, and the coverage by kG fields doubles as a strong mesogranular network is formed \citep{2014ApJ...789..132R}.

Our results seem to reveal a hierarchical picture for the evolution of quiet Sun magnetic fields: flux emerges preferentially near the edges of  granules in the form of small-scale magnetic loops \citep{2007ApJ...666L.137C,2009ApJ...700.1391M,2010ApJ...723L.149D}; the footpoints are first swept to nearby intergranular lanes unless they cancel out before \citep{2007ApJ...666L.137C,2014ApJ...789....6R}, and then advected to mesogranular lanes, particularly to the junctions where long-lived downdrafts are formed. The concentration of weak fields in the mesogranular vertices gives rise to internetwork magnetic elements. If these magnetic elements are strong (long-lived) enough, supergranular flows can drag them along \citep{2008ApJ...684.1469D,2012ApJ...758L..38O}, until they reach the supergranular boundaries and eventually contribute to the magnetic network flux \citep{2014ApJ...797...49G}. Interestingly, this qualitative scenario may explain why most of the newly detected internetwork flux is not in the form of bipolar elements, but instead seems to be unipolar \citep{2008ApJ...674..520L,2015PhDT.......G,2016arXiv160808499A}. Unipolar appearances are thought to be the result of the coalescence of previously unobservable, very weak flux into stronger features, as first pointed out by \citet{2008ApJ...674..520L}. The concentration of weak fields in sinks provides a plausible mechanism whereby this
can occur. 

\acknowledgments

We thank M. J. Mart\'{i}nez Gonz\'{a}lez for providing the locations of magnetic loops shown in Figure \ref{fig1}. We also thank the referee for constructive suggestions that contribute to improve the article. The work by I. S. R. has been funded by the Basque Government under a grant from Programa Predoctoral de Formaci\'{o}n de Personal Investigador del Departamento de Educaci\'{o}n, Universidades e Investigaci\'{o}n.
This work has been partially funded by the Spanish Ministerio de Econom\'{\i}a y Competitividad, 
through Projects No. ESP2013-47349-C6-1-R  and ESP2014-56169-C6-1-R, including a percentage from European FEDER funds.
The German contribution has been funded by the Bundesministerium f\"ur Wirtschaft und Technologie through Deutsches Zentrum f\"ur Luft- und Raumfahrt e.V. (DLR), grant number 50 OU 0401, and by the Innovationsfond of the President of the Max Planck Society (MPG). 
This work was partly supported by the BK21 plus program through the National Research Foundation (NRF) funded by the Ministry of Education of Korea.

\end{document}